\documentclass[english,showpacs,nofootinbib,citesort]{revtex4}
\usepackage[T1]{fontenc}
\usepackage[latin1]{inputenc}
\usepackage{graphicx}

\makeatletter

\usepackage{color}
\usepackage{graphicx}

\usepackage{babel}
\makeatother
\begin{document}
\newcommand\gsim{\mathrel{\rlap{\raise.4ex\hbox{$>$}} {\lower.6ex\hbox{$\sim$}}}}   \newcommand\lsim{\mathrel{\rlap{\raise.4ex\hbox{$<$}} {\lower.6ex\hbox{$\sim$}}}}

\begin{flushright} hep-ph/0506309 
\end{flushright}

\title{Off-shell scattering amplitudes in the double-logarithmic approximation}

\author{B.I.~Ermolaev}

\affiliation{Ioffe Physico-Technical Institute, 194021 St.Petersburg, Russia}

\author{M.~Greco}

\affiliation{Department of Physics and INFN, University Rome III, Rome, Italy}

\author{F.~Olness}

\affiliation{Department of Physics, Southern Methodist University, Dallas, TX
75275-0175 USA}

\author{S.I.~Troyan}

\affiliation{St.Petersburg Institute of Nuclear Physics, 188300 Gatchina, Russia}

\begin{abstract}
When scattering amplitudes are calculated in the double-logarithmic
approximation, it is possible to relate the double-logarithmic on-shell
and off-shell amplitudes. Explicit relations are obtained for scattering
amplitudes in QED, QCD, and the ElectroWeak Standard Model. The off-shell
amplitudes are considered in the hard and the Regge kinematic limits.
We compare our results in both the Feynman and Coulomb gauges. 
\end{abstract}


\maketitle

\section{Introduction}

Because off-shell scattering amplitudes are not gauge-invariant quantities,
they have not been studied as extensively as their on-shell counterparts.
Nevertheless, off-shell scattering amplitudes are important to calculate
as they arise in many well-known cases, such as when the process of
interest can be divided into sub-processes. One example is the $e^{+}e^{-}$
annihilation into off-shell unstable particles (a quark pair, or a
$W^{+}W^{-}$- pair) followed by the decay of the off-shell states.
Other examples include the Bethe-Salpeter equation which relates off-shell
amplitudes to on-shell amplitudes, and the DGLAP\cite{dglap} evolution
equations which operate with off-shell amplitudes. The Deep-Inelastic
Scattering (DIS) structure functions provide yet another example when
the initial parton is not assumed to be on the mass shell.

In general, the expressions for the on-shell and off-shell scattering
amplitudes are quite different and should be calculated independently.
There exist certain cases where it is possible to relate off-shell
and on-shell amplitudes; however, even if we limit our scope to the
the double logarithmic approximation (DLA), there are many cases where
the on-shell amplitude cannot be obtained by a simple limit of the
off-shell amplitude.

In order to demonstrate the relation between the off-shell and on-shell
amplitudes, let us consider the well-known example of the Sudakov
form factor in QED.\cite{sud} When the virtualities of the initial
$(p_{1}^{2})$ and the final $(p_{2}^{2})$ electrons are much greater
than the electron mass squared $(|p_{1,2}^{2}|\gg m^{2})$, the expression
for the off-shell Sudakov form-factor $S(q^{2})$ of electron in QED
is: \begin{equation}
S(q^{2})_{off}=\exp[{-(\alpha/2\pi)\ln(-q^{2}/|p_{1}^{2}|)\ln(-q^{2}/|p_{2}^{2}|)}]\label{sudoff}\end{equation}
 where $q$ is the momentum transfer: $q=p_{2}-p_{1}$. We note that
Eq.~(\ref{sudoff}) accounts for all DL contributions to the first
part $(\sim\gamma_{\mu})$ of the photon-electron vertex%
\footnote{The expressions for the DL asymptotics of the second part ($\sim\sigma_{\mu\nu}$)
of the photon-fermion vertex are obtained in Ref.~\cite{et}%
} in the limit $|q^{2}|\gg|p_{1,2}^{2}|$. Conversely, when the electron
is on-shell before and after the scattering process, the Sudakov form-factor
is \begin{equation}
S(q^{2})_{on}=\exp[-(\alpha/4\pi)\ln^{2}(-q^{2}/m^{2})]\,.\label{sudon}\end{equation}
 The minus sign in the logarithmic terms $\ln(-q^{2})$ of Eqs.~(\ref{sudoff},\ref{sudon})
is related to the analyticity in the $q^{2}$-plane; indeed, the Sudakov
form factor should not have an imaginary part when $q^{2}<0$. To
simplify the notation, we will drop this minus sign, together with
the modulus for $|p_{1,2}^{2}|$, throughout the paper.

Obviously, Eq.~(\ref{sudon}) cannot be obtained from Eq.~(\ref{sudoff})
by taking the limit $p_{1}^{2}=p_{2}^{2}=m^{2}$. Hence, the on-shell
and off-shell Sudakov form-factors cannot be related by a simple analytic
continuation of the mass scales. This result stems from the fact that
the DL contributions to the Sudakov form-factors come from infrared
singularities in the integration over the virtual photon momenta;
the separate mass scales $m^{2}$ and $p_{1,2}^{2}$ regulate these
singularities.

In the one-loop approximation, the general expression for the DL contribution
to the Sudakov form-factors can be written as: \begin{equation}
W=-(\alpha/4\pi)\Big[\ln^{2}(q^{2}/m^{2})-\ln^{2}(p_{1}^{2}/m^{2})-\ln^{2}(p_{2}^{2}/m^{2})+\Theta(p_{1}^{2}p_{2}^{2}-m^{2}q^{2})\ln^{2}(m^{2}q^{2}/p_{1}^{2}p_{2}^{2})\Big]\label{oneloop}\end{equation}
 where the value of the infrared cut-off $\mu$ is chosen as ($\mu\approx m$).
Summing the higher loop contributions for $W$ leads to the exponentiation
of the one-loop result. The individual terms in the squared brackets
in Eq.~(\ref{oneloop}) are obtained from the integration over separate
kinematic regions.

The calculation in DLA is especially simple when the Sudakov parametrization
is used for the soft momenta $k_{\mu}$:

\begin{equation}
k_{\mu}=\alpha\,{p'}_{2\,\mu}+\beta\,{p'}_{1\,\mu}+k_{\perp\,\mu}\,,\label{sudvar}\end{equation}
 with ${p'}_{1\,\mu}=p_{1\,\mu}-(p_{1}^{2}/q^{2})p_{2\,\mu}$ and
${p'}_{2\,\mu}=p_{2\,\mu}-(p_{2}^{2}/q^{2})p_{1\,\mu}$ so that ${p'}_{1}^{2}={p'}_{2}^{2}=0,~2{p'}_{1}\,{p'}_{2}=q^{2}$.
Additionally, $\alpha$ and $\beta$ are parameters, and $k_{\perp~\mu}$
is the component of $k_{\mu}$ orthogonal to both $p_{1}$ and $p_{2}$.

Making use of the Sudakov parameterization, we can trace the origin
of the separate terms for the one-loop form factor of Eq.~(\ref{oneloop}).
The first term {[}$\ln^{2}(q^{2}/m^{2})${]} arises from the integration
over the region $1\gg\alpha,~\beta\gg m^{2}/q^{2}$ and $\alpha\,\beta\gg m^{2}/q^{2}$
for the case of an on-shell electron. When the initial (final) electron
is off-shell, the condition $\alpha\gg p_{1}^{2}/q^{2}$ ($\beta\gg p_{2}^{2}/q^{2}$)
applies (provided the virtualities of the electron are so small that
$p_{1}^{2}\, p_{2}^{2}<m^{2}q^{2}$) and we obtain the second {[}$-\ln^{2}(p_{1}^{2}/m^{2})${]}
and the third {[}$-\ln^{2}(p_{2}^{2}/m^{2})${]} DL terms of Eq.~(\ref{oneloop}).
Finally, when $p_{1}^{2}\, p_{2}^{2}>m^{2}q^{2}$, the integration
region is $1\gg\alpha\gg p_{1}^{2}/q^{2},~1\gg\beta\gg p_{2}^{2}/q^{2}$,
and this leads to the last term {[}$\Theta(p_{1}^{2}\, p_{2}^{2}-m^{2}q^{2})\ln^{2}(m^{2}q^{2}/p_{1}^{2}\, p_{2}^{2})${]}
of Eq.~(\ref{oneloop}).

When considering only the first three terms of Eq.~(\ref{oneloop}),
the off-shell $W$ and on-shell $W$ are related analytically. One
can easily take the limit $p_{1}^{2},p_{2}^{2}\rightarrow m^{2}$
in the off-shell $W$ to obtain the on-shell $W$. However, the last
term {[}$\Theta(p_{1}^{2}p_{2}^{2}-m^{2}q^{2})\ln^{2}(m^{2}q^{2}/p_{1}^{2}p_{2}^{2})${]}
of Eq.~(\ref{oneloop}) changes this situation and results in a non-trivial
relation between the on-shell and off-shell form factors.

Beyond QED, problems relating $S(q^{2})_{on}$ and $S(q^{2})_{off}$
also exist in QCD and in the EW Standard Model. Additionally, when
calculating DL scattering amplitudes in the EW Standard Model at energies
$\gg$ 100~GeV$^{2}$, the situation becomes more involved because,
in addition to the virtual photon exchanges, one must include $W$
and $Z$ exchanges. In contrast to QED, the logarithmic contributions
from the $W$ and $Z$ bosons are regulated by their masses ($M_{W}$
and $M_{Z}$); therefore, one should not introduce an infrared cut-off
for them.

To simplify our notation, we introduce the mass scale $M$

\begin{equation}
M\approx M_{Z}\approx M_{W}\,,\label{M}\end{equation}
 which we will use to treat the $W$ and $Z$ exchange processes in
a more symmetric manner. We stress that introducing $M$ is purely
a technical point, which is useful for performing an all-order summations
of DL contributions in the Standard Model.

Most of the available DL calculations of EW on-shell amplitudes have
been performed in the one-loop and two-loop approximation in the EW
couplings,\cite{e,bw} An all-orders DL summation for the EW on-shell
amplitudes is also well-studied for the case of the hard kinematics
where the total resummation of DL contributions exponentiates\cite{flmm,k}.
The more complicated case of the Regge kinematics was studied in detail
in Refs.~\cite{b}.

In the present paper, we relate the off-shell and on-shell double-logarithmic
scattering amplitudes in QED, QCD, and the EW Standard Model. We obtain
explicit expressions for the off-shell double-logarithmic scattering
amplitudes in the Feynman gauge (as the off-shell amplitudes are gauge-dependent),
and we also discuss the Coulomb gauge.

The organization of the paper is as follows. In Sect.~2 we relate
the double-logarithmic off-shell and on-\textcolor{black}{shell amplitudes
in QED in the hard kinematic region. I}n Sec.~3 we consider more
involved examples of the off-shell $2\rightarrow2$ scattering amplitudes
where both virtual photons and $W,Z$ bosons are accounted to all
orders in the electroweak couplings. In Sect.~4 we generalize the
technique of Sect.~2 to the case of the forward Regge kinematics,
and obtain explicit expressions for the DIS non-singlet structure
function $g_{1}$ with the initial quark being off-shell. Specifically,
we examine the effect of the initial quark virtuality on the small-$x$
asymptotics of $g_{1}$. Finally, in Sect.~5 we present our conclusions.

\section{Relations between on-shell and off-shell amplitudes in QED}

\begin{figure}[t]
\includegraphics{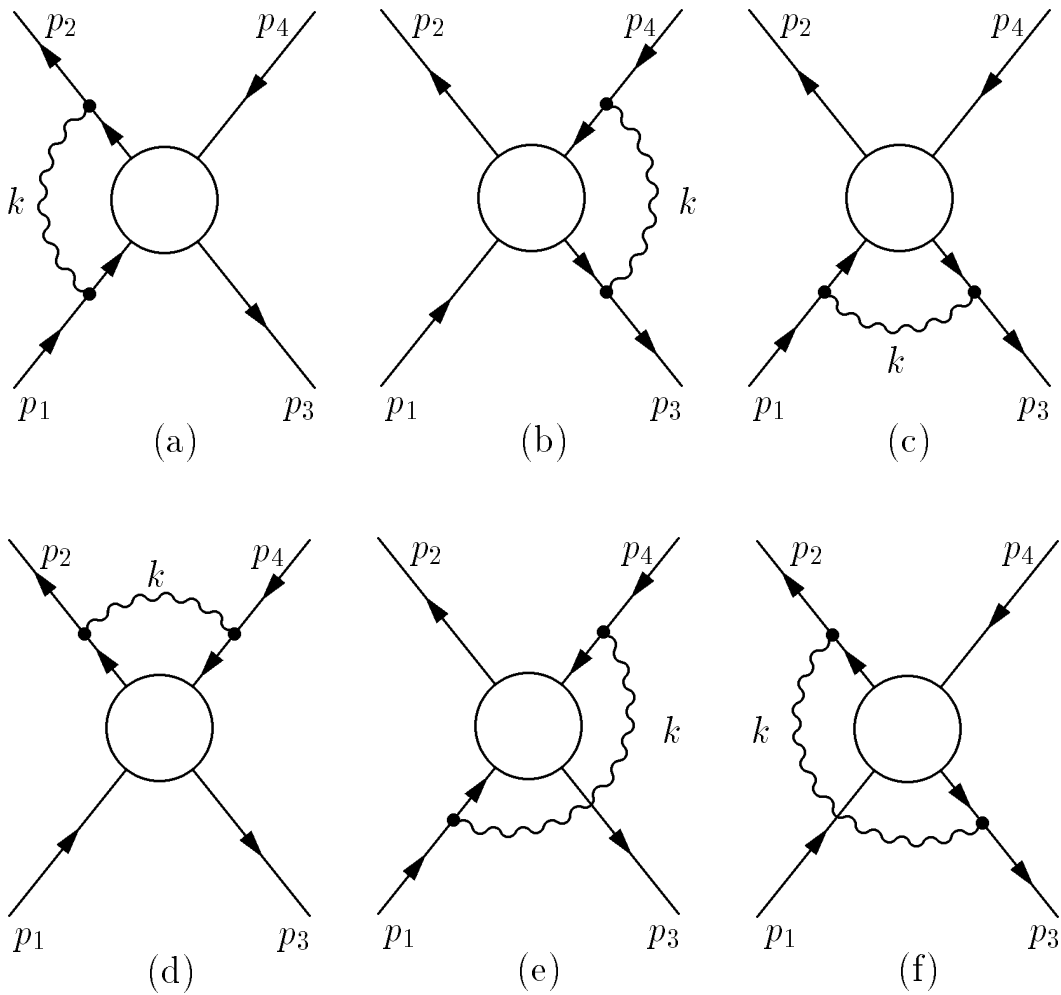}

\caption{Diagrams showing the factorization of the softest photon with momentum
$k$ when the Feynman gauge is used.}
\end{figure}

\begin{figure}[t]
\includegraphics{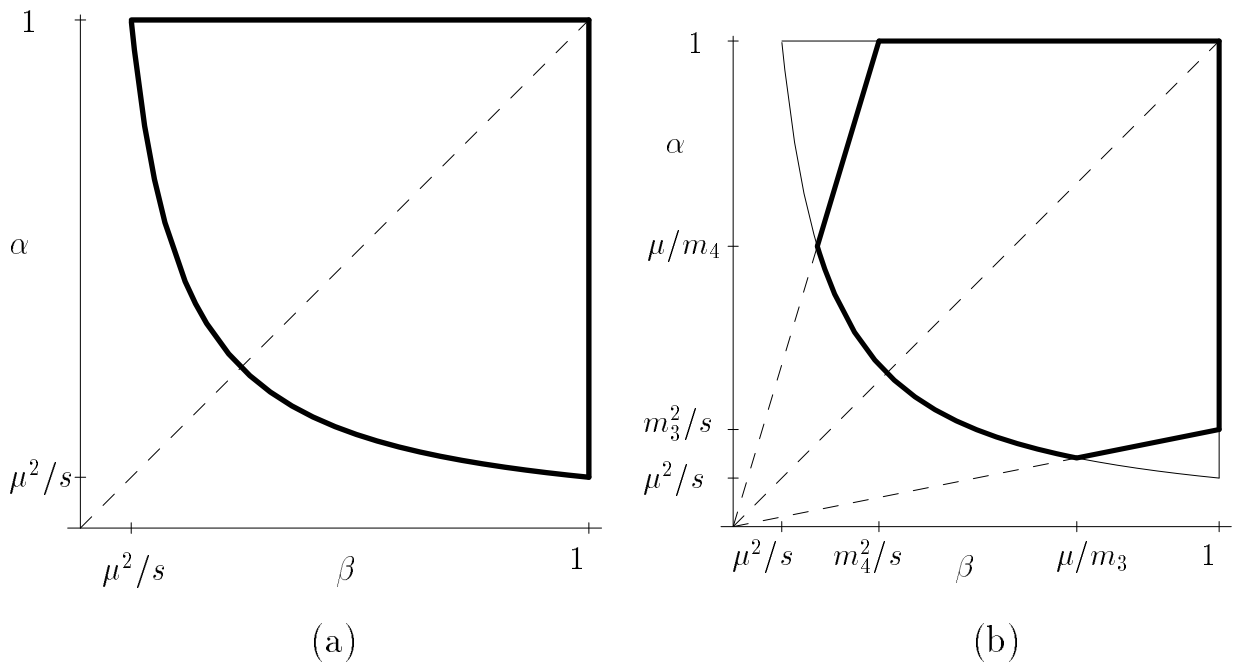}

\caption{The integration region over the softest photon momentum when all
external particles are on-shell and the infrared cut-off $\mu$ is
chosen (a) as in Eq.~(\ref{mubig}), and (b) as in Eq.~(\ref{musmall}).}
\end{figure}

In the present section, we obtain a general prescription for relating
off-shell and on-shell scattering amplitudes in the DLA for the (simplest)
case of QED. We consider a $2\rightarrow2$ process in QED, where
the initial and final particles are fermions. Although we derive our
results for a general process, we will use the $e^{+}e^{-}$ annihilation
into quark or lepton pairs {[}e.g., $e^{-}(p_{1})e^{+}(p_{2})\rightarrow q^{-}(p_{3})q^{+}(p_{4})${]}
as our {}``model'' process; the final formula however are valid
for any channel. For this $2\rightarrow2$ process, we work in the
hard kinematic limit where:

\begin{equation}
s=(p_{1}+p_{2})^{2}\sim-t=-(p_{3}-p_{1})^{2}\sim-u=(p_{1}-p_{4})^{2}\gg|p_{1,2}^{2}|,|p_{3,4}^{2}|\,.\label{hard}\end{equation}

We will use the Infrared Evolution Equations (IREE) to resum the DL
contributions.%
\footnote{For a recent alternate approach, see Refs.~\cite{k,ciaf,kur}.%
} The IREE method has successfully been used to calculate the scattering
amplitudes of various processes in QED, QCD and the EW Standard Model.%
\footnote{See Refs.~\cite{et,flmm,b,kl,efl,egt1}, and the references therein. %
} The first step is to use the IREE to factorize the DL contributions
from the {}``softest'' virtual particle, which we define to be the
particle with the minimal transverse momentum $k_{\perp}$.

For QED scattering amplitudes in the hard kinematic limit Eq.~(\ref{hard}),
the softest particles must be photons. Their DL contributions can
be factorized according to the Gribov bremsstrahlung theorem.%
\footnote{The generalization of Gribov bremsstrahlung theorem\cite{g} to QCD
discussed in Refs.~\cite{efl,ce,kl}.%
} The integration region over the softest photon momentum contains
an infrared singularity, and therefore we must introduce an infrared
cut-off, $\mu$. We impose this cut-off in the transverse momentum
space by restricting the transverse momenta of all virtual particles
to be greater than $\mu$. This constraint modifies the boundaries
of the integration region, and results in relations between the cut-off
scale $\mu$ and the masses (virtualities) of the external particles.

\subsection{The case of the on-shell amplitude
$A(s,\mu^{2})$ with  massless  external particles}

We begin with the well-know case where all the external particles
are on-shell:

\begin{equation}
p_{1,2}^{2}=m_{1,2}^{2},~~p_{3,4}^{2}=m_{3,4}^{2}\,.\label{onshellp}\end{equation}
 For simplicity, we take the case where the fermion masses are negligible:\begin{equation}
\mu^{2}>m_{1,2}^{2},~m_{3,4}^{2}\,.\label{mubig}\end{equation}
 For this case, in the hard kinematic region the amplitude $A$ only
depends on $s$ and $\mu^{2}$, and we can easily obtain the IREE
for this process. The RHS of the IREE includes the Born term, $A^{Born}$,
and the contributions of the graphs depicted in Fig.~1. These graphs
represent all the possible ways to factorize the softest photon, providing
the Feynman gauge is used. Using the standard Feynman rules, we obtain
at the following IREE:

\begin{equation}
A(s,\mu^{2})=A^{Born}-2\lambda Q^{2}\int_{\mu^{2}}^{s}\frac{dk_{\perp}^{2}}{k_{\perp}^{2}}\ln(s/k_{\perp}^{2})A(s,k_{\perp}^{2})\,,\label{inton}\end{equation}
 where $\lambda=\alpha/8\pi$, and $Q$ is the modulus of the electric
charge of the fermions. In Eq.~(\ref{inton}) recall that $k_{\perp}$
plays the role of a new cut-off for any other virtual particles inside
the blobs in Fig.~1,

The logarithmic factor in the integrand of Eq.~(\ref{inton}) arises
from the integration over the longitudinal Sudakov variable $\beta$
{[}see Eq.~(\ref{sudvar}){]} of the factorized photon in the region
$\mu^{2}\ll k_{\perp}^{2}\ll s\beta\ll s$. The integration region
over $\beta$ and $k_{\perp}^{2}$ is especially simple in terms of
the variables $\beta$ and $\alpha\equiv k_{\perp}^{2}/s\beta$, and
this region is depicted in Fig.~2a. Differentiation with respect
to $\ln\mu^{2}$ converts the integral equation Eq.~(\ref{inton})
into a differential one: \begin{equation}
\partial A(x)/\partial x=-2\lambda Q^{2}xA(x)\label{difon}\end{equation}
 with the solution \begin{equation}
A(s,\mu^{2})=A^{B}\exp(-\lambda Q^{2}x^{2})\,.\label{solon}\end{equation}
 Here, we have introduced the following notations: $x=\ln(s/\mu^{2})$,
$\lambda=\alpha/(8\pi)$, and $Q^{2}$ is defined to be:\begin{equation}
Q^{2}=Q_{1}^{2}+Q_{2}^{2}+Q_{3}^{2}+Q_{4}^{2}\,,\label{q}\end{equation}
 with $Q_{i}~(i=1,2,3,4)$ being the modulus of the electric charge
of the initial and final particles in units of $e=\sqrt{4\pi\alpha}$.
As we are using the Feynman gauge, the DL contributions from the graphs
depicted in Fig.~1 yield another constraint for $Q^{2}$: \begin{equation}
Q^{2}=Z_{a}+Z_{b}+Z_{c}+Z_{d}-Z_{e}-Z_{f}\label{q2}\end{equation}
 where the subscripts $a,..,f$ refer to the graphs $a,..,f$ in Fig.~1.
Furthermore, we have the additional relations \begin{eqnarray}
Z_{a} & \equiv & 2Q_{1}Q_{2}\nonumber \\
Z_{b} & \equiv & 2Q_{3}Q_{4}\nonumber \\
Z_{c} & \equiv & 2Q_{1}Q_{3}\nonumber \\
Z_{d} & \equiv & 2Q_{2}Q_{4}\label{qi}\\
Z_{e} & \equiv & 2Q_{1}Q_{4}\nonumber \\
Z_{f} & \equiv & 2Q_{2}Q_{3}.\nonumber \end{eqnarray}
 Finally, making use of electric charge conservation%
\footnote{In the explicit formula of this paper we refer to the annihilation
channel, although the final results are true in any channel.%
}\begin{equation}
Q_{1}-Q_{2}-Q_{3}+Q_{4}=0\,.\label{charge}\end{equation}
 we can rewrite $Q^{2}$ in the form of Eq.~(\ref{q}).

\subsection{The case of the on-shell amplitude
$\tilde{A}(x,y_{1},y_{2})$ with  massive external particles}

\begin{figure}[t]
\includegraphics{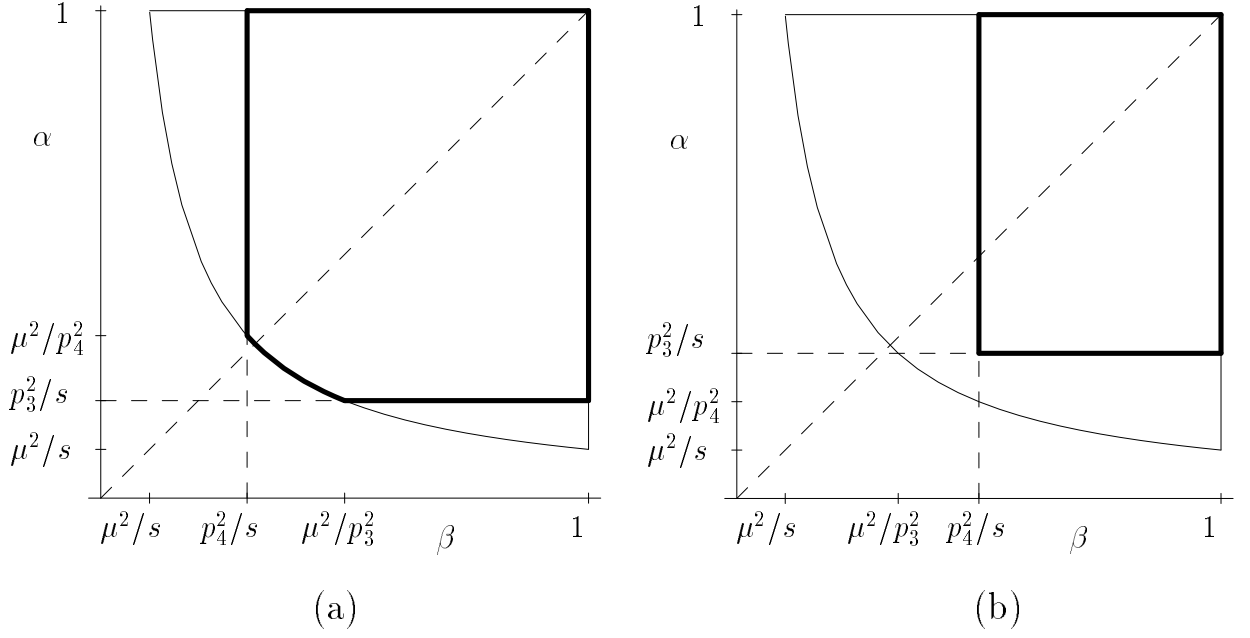}
\caption{The integration region over the softest photon momentum when the
final particles are (a) moderately-virtual and (b) deeply-virtual.}
\end{figure}

When $\mu$ is less than some of the masses of the external particles,
the situation becomes more complex. We focus here on the specific
case where:\begin{equation}
m_{1,2}<\mu<m_{3,4}.\label{musmall}\end{equation}
 In this case, the scattering amplitude $\tilde{A}$ depends on $m_{3,4}^{2}$
in addition to $s$ and $\mu^{2}$. The integration region in the
$\{\alpha,\beta\}$ plane%
\footnote{Recall $\alpha\equiv k_{\perp}^{2}/s\beta$, and $s\beta$ is the
longitudinal momentum of the softest photon.%
} is depicted in Fig.~2b; note how this differs from the integration
region for Eq.~(\ref{inton}) shown in Fig.~2a. Performing the $\beta$
integration over the region of Fig.~2b, and differentiating the remaining
integral with respect to $\mu$, we arrive at the new IREE in the
differential form:

\begin{equation}
\partial\tilde{A}/\partial x+\partial\tilde{A}/\partial y_{1}+\partial\tilde{A}/\partial y_{2}=-2\lambda(Q^{2}x-Z_{A}y_{1}/2-Z_{B}y_{2}/2)\tilde{A},\label{difmusmall}\end{equation}
 with $y_{1,2}=\ln(m_{3,4}^{2}/\mu^{2})$, $Z_{A}=(Z_{b}+Z_{c}-Z_{f})=2Q_{3}^{2}$,
and $Z_{B}=Z_{b}+Z_{d}-Z_{e}=2Q_{4}^{2}$. The general solution to
Eq.~(\ref{difmusmall}) is

\begin{equation}
\tilde{A}(x,y_{1},y_{2})=\tilde{\Phi}(x-y_{2},~y_{1}-y_{2})\exp\big[-\lambda y_{2}\big(2Q^{2}x-2Z_{A}y_{1}+(-Q^{2}+Z_{A}-Z_{B})y_{2}\big)\big].\label{gensolmusmall}\end{equation}

For $y_{2}=\ln(m_{4}^{2}/\mu^{2})=0$ (i.e., $m_{4}^{2}=\mu^{2}$)
there should be an obvious matching condition\begin{equation}
\tilde{A}(x,y_{1},y_{2})|_{y_{2}=0}=\tilde{A'}(x,y_{1}),\label{matchmusmall}\end{equation}
 where $\tilde{A'}(x,y_{1})$ is the amplitude of the same process,
but for the case where $m_{1,2}<m_{4}\ \,\lsim\,\mu<m_{3}$.
This matching allows us to fix the unknown function $\tilde{\Phi}(x-y_{2},y_{1}-y_{2})$,
providing the amplitude $\tilde{A'}(x,y_{1})$ is known. Taking the
$y_{2}=0$ limit of Eq.~(\ref{difmusmall}), it is easy to see that
$\tilde{A'}(x,y_{1})$ obeys the following IREE: \begin{equation}
\partial\tilde{A'}/\partial x+\partial\tilde{A'}/\partial y_{1}=-2\lambda(Q^{2}x-Z_{A}y_{1}/2)\tilde{A}\label{difmusmall1}\end{equation}
 with the general solution \begin{equation}
\tilde{A'}(x,y_{1})=\tilde{\Phi'}(x-y_{1})\exp\big[-\lambda y_{1}\big(2Q^{2}x-2Z_{A}y_{1}+(-Q^{2}-Z_{A})y_{1}\big)\big]\,.\label{gensolmusmall1}\end{equation}
 In order to determine the function $\tilde{\Phi'}(x-y_{1})$, we
again apply the matching condition for $y_{1}=\ln(m_{3}^{2}/\mu^{2})=0$
($m_{3}^{2}=\mu^{2}$)\begin{equation}
\tilde{A'}(x,y_{1})|_{y_{1}=0}=A(x)\quad,\label{matchmusmallon}\end{equation}
 where the amplitude $A(x)$ is defined in Eq.~(\ref{solon}). The
matching condition of Eq.~(\ref{matchmusmallon}) immediately leads
to the solution for $\tilde{A'}(x,y_{1})$:\begin{equation}
\tilde{A'}(x,y_{1})=A(x-y_{1})\exp\big[-\lambda y_{1}\big(2Q^{2}x-2Z_{A}y_{1}+(-Q^{2}-Z_{A})y_{1}\big)\big]\,.\label{solsmallmu1}\end{equation}
 Inserting this result in Eq.~(\ref{matchmusmall}), we can obtain
$\tilde{A}(x,y_{1},y_{2})$ in terms of $\tilde{A'}(x,y_{1})$\begin{equation}
\tilde{A}(x,y_{1},y_{2})=\tilde{A'}(x-y_{2},y_{1}-y_{2})\exp\big[-\lambda y_{2}\big(2Q^{2}x-2Z_{A}y_{1}+(-Q^{2}+Z_{A}-Z_{B})y_{2}\big)\big]\,.\label{solmusmall}\end{equation}
 Finally, after performing all the substitutions, we find:\begin{equation}
\tilde{A}(x,y_{1},y_{2})=A^{Born}\exp\big[-\lambda\big(Q^{2}x^{2}-Z_{A}y_{1}^{2}/2-Z_{B}y_{2}^{2}/2\big)\big]\,.\label{solmusmallexpl}\end{equation}
 In general, one can use Eq.~(\ref{matchmusmall}) when $m_{3}>m_{4}$.

\subsection{The case of 
$A_{1}(s,\mu^{2},p_{3}^{2})$ with one off-shell final-state particle}

Let us consider now the situation when one of the final-state particles
is off-shell: \begin{equation}
s\gg p_{3}^{2}\gg\mu^{2}\,.\label{oneoff}\end{equation}
 We denote the amplitude for this case as $A_{1}(s,\mu^{2},p_{3}^{2})$.
Let us replace this by: In the DLA, this amplitude does not depend
on $m_{3}$; therefore $m_{3}$ can be dropped from Eq.~(\ref{mubig})
and Eq.~(\ref{musmall}) relating the value of the cut-off and the
fermion masses. However, in order to use the matching between the
off-shell and on-shell amplitudes in the simplest manner, we will
assume that $\mu$ obeys Eq.~(\ref{mubig}). In contrast to the case
of the on-shell amplitudes, the softest photon can now be factorized
out of $A_{1}(s,\mu^{2},p_{3}^{2})$ when $\mu^{2}\ll k_{\perp}^{2}\ll p_{3}^{2}$.
As was observed in the previous case, the $\beta-$integration region
of the longitudinal momentum $s\beta$ of the softest photon momentum
is now different for each individual graph in Fig.~1. Specifically,
when the propagator of the photon connects to an on-shell external
line, the integration region (depicted in Fig.~2a) yields the factor
$\ln(s/k_{\perp}^{2})$. Conversely, when the off-shell line with
momentum $p_{3}$ is involved, the integration region (depicted in
Fig.~3a) yields instead the factor $\ln(s/p_{3}^{2})$. This leads
to the following integral form for the IREE:

\begin{eqnarray}
A_{1}(s/\mu^{2},{p'}^{2}/\mu^{2}) & = & A_{1}^{Born}-2\lambda\Big[Z_{C}\int_{\mu^{2}}^{{p'}^{2}}\frac{dk_{\perp}^{2}}{k_{\perp}^{2}}\ln(s/k_{\perp}^{2})A_{1}(s/k_{\perp}^{2},p_{3}^{2}/k_{\perp}^{2})\label{intoffone}\\
 & = & +Z_{A}\int_{\mu^{2}}^{{p'}^{2}}\frac{dk_{\perp}^{2}}{k_{\perp}^{2}}\ln(s/p_{3}^{2})A_{1}(s/k_{\perp}^{2},p_{3}^{2}/k_{\perp}^{2})\Big]\nonumber \end{eqnarray}
 Here, we have defined $Z_{C}=Q^{2}-Z_{A}$ where $Z_{A}$ was introduced
in Eq.~(\ref{difmusmall}). Differentiating Eq.~(\ref{intoffone})
with respect to $\mu^{2}$, we obtain a partial differential equation
for $A_{1}(s,\mu^{2},p_{3}^{2})$

\begin{equation}
\partial A_{1}/\partial x+\partial A_{1}/\partial z_{1}=-2\lambda(Q^{2}x-Z_{A}z_{1})A_{1}\label{difoffone}\end{equation}
 where we have introduced the new variable $z_{1}=\ln(p_{3}^{2}/\mu^{2})$.
The solution to Eq.~(\ref{difoffone}) is given by

\begin{equation}
A_{1}(x,z_{1})=\Phi(x-z_{1})\exp\big[-2\lambda Q^{2}(x-z_{1})z_{1}+\lambda Z_{A}z_{1}^{2})\big]\label{gensoloffone}\end{equation}
 where the function $\Phi$ is, as yet, unknown. In order to determine
$\Phi$, we use the matching conditions at $z_{1}=\ln(p_{3}^{2}/\mu^{2})=0$
(i.e., $p_{3}^{2}=\mu^{2}$)

\begin{equation}
A_{1}(x,z_{1})|_{z_{1}=0}=A(x).\label{matchone}\end{equation}
 With this relation, we then obtain

\begin{equation}
A_{1}(x,z_{1})=A(x-z_{1})\exp\big[-2\lambda Q^{2}(x-z_{1})z_{1}-\lambda Z_{C}z_{1}^{2})\big]\label{soloffone}\end{equation}
 which relates the off-shell amplitude $A_{1}$ to the on-shell amplitude
$A$. Using Eq.~(\ref{solon}), we obtain an explicit expression
for $A_{1}(x,z_{1})$: \begin{equation}
A_{1}(x,z_{1})=A^{Born}\exp\big[-\lambda(Q^{2}x^{2}-Z_{A}z_{1}^{2})\big].\label{solone}\end{equation}

\subsection{The case of 
 $A_{2}(s,p_{3}^{2},p_{4}^{2},\mu^{2})$ 
and
$\widetilde{A}_{2}(x,y_{1},y_{2})$ with 
two off-shell final-state particles}

Let us consider now the amplitude $A_{2}(s,p_{3}^{2},p_{4}^{2},\mu^{2})$
where both the final-state particles are off-shell. The amplitude
$A_{2}(s,p_{3}^{2},p_{4}^{2},\mu^{2})$ will obey different IREE depending
on the value of the virtualities; we will address these in turn.

\subsubsection{Moderately-Virtual Kinematics: 
$A_{2}(s,p_{3}^{2},p_{4}^{2},\mu^{2})$}

We first consider the case where $p_{3,4}^{2}>\mu^{2}$ but with the
condition\begin{equation}
p_{3}^{2}\, p_{4}^{2}<s\mu^{2}\,.\label{smallvirt}\end{equation}
 We will refer to this as the moderately-virtual case. The IREE for
$A_{2}(x,z_{1},z_{2})$ is similar to Eq.~(\ref{difoffone}): \begin{equation}
\partial A_{2}/\partial x+\partial A_{2}/\partial z_{1}+\partial A_{2}/\partial z_{2}=-2\lambda(Q^{2}x-Z_{A}z_{1}-Z_{B}z_{2})A_{2},\label{difoffsmall}\end{equation}
 where we define $z_{1,2}\equiv\ln(p_{3,4}^{2}/\mu^{2})$, and $Z_{A,B}$
is introduced in Eq.~(\ref{difmusmall}). Again, we solve this equation
and using the matching conditions at $z_{2}=0$ (i.e., $p_{3}^{2}=\mu^{2}$)

\begin{equation}
A_{2}(x,z_{1},z_{2})|_{z_{2}=0}=A_{1}(x,z_{1})\,,\label{matchsmall}\end{equation}
 and eventually obtain \begin{equation}
A_{2}(x,z_{1},z_{2})=A^{Born}\exp\Big[-\lambda\big(Q^{2}x^{2}-Z_{A}z_{1}^{2}-Z_{B}z_{2}^{2}\big)\Big].\label{solsmall}\end{equation}

\subsubsection{Deeply-Virtual Kinematics: 
$\widetilde{A}_{2}(x,y_{1},y_{2})$}

Finally, let us consider the case of the deeply-virtual kinematics\begin{equation}
p_{3}^{2}\, p_{4}^{2}>s\mu^{2}\,.\label{bigvirt}\end{equation}
 We denote the deeply-virtual amplitude as $\widetilde{A}_{2}(x,y_{1},y_{2})$.
As the integration in graph (b) of Fig.~1 does not depend on $\mu$,
the IREE is similar to that of Eq.~(\ref{difoffsmall}), with the
exception of the contribution proportional to $Z_{b}$: \begin{equation}
\partial\tilde{A}_{2}/\partial x+\partial\tilde{A}_{2}/\partial z_{1}+\partial\tilde{A}_{2}/\partial z_{2}=-2\lambda(\tilde{Q}^{2}x-\tilde{Z}_{A}z_{1}-\tilde{Z}_{B}z_{2})\tilde{A}_{2}.\label{difoffbig}\end{equation}
 Here, the replacement $\tilde{Q}^{2}=Q^{2}-Z_{b},~\tilde{Z}_{A,B}=Z_{A,B}-Z_{b}$
should be made because the integration of graph (b) in Fig.~1 is
performed over the region depicted in Fig.~3b. The solution of Eq.~(\ref{difoffbig})
is \begin{equation}
\tilde{A}_{2}(x,y_{1},y_{2})=\tilde{\Phi}_{2}(x-z_{1},x-z_{2})\exp\Big[-\lambda\big(\tilde{Q}^{2}x^{2}-\tilde{Z}_{A}z_{1}^{2}-\tilde{Z}_{B}z_{2}^{2}\big)\Big]\,.\label{gensolbig}\end{equation}
 We determine the unknown function $\tilde{\Phi}_{2}$ using the matching
condition at $p_{3}^{2}p_{4}^{2}=s\mu^{2}$ (i.e., when $x=z_{1}+z_{2}$),
and obtain:

\begin{equation}
\tilde{A}_{2}(x,y_{1},y_{2})|_{x=z_{1}+z_{2}}=A_{2}(x,z_{1},z_{2})\,.\label{matchbs}\end{equation}
 For the function $\tilde{\Phi}_{2}$ we obtain\begin{equation}
\tilde{\Phi}(x-z_{1},x-z_{2})=\exp\Big[-2\lambda Z_{b}\big(x-z_{1})(x-z_{2}\big)\Big]\,,\label{matchbig2}\end{equation}
 and therefore the deeply-virtual amplitude $\tilde{A}_{2}(x,z_{1},z_{2})$
is given by:\begin{eqnarray}
\tilde{A}_{2}(x,z_{1},z_{2}) & = & A\exp{\Big[-\lambda[\tilde{Q}^{2}x^{2}-\tilde{Z}_{A}z_{1}^{2}-\tilde{Z}_{B}z_{2}^{2}+2Z_{b}(x-z_{1})(x-z_{2})]\Big]}\label{solbig}\\
 & = & A^{Born}\exp{\Big[-\lambda[Q^{2}x^{2}-Z_{A}z_{1}^{2}-Z_{B}z_{2}^{2}+Z_{b}(x-z_{1}-z_{2})^{2}]\Big]}.\nonumber \end{eqnarray}
 The last term in the exponent of Eq.~(\ref{solbig}) contains the
factor $Z_{b}$ defined by:

\begin{equation}
Z_{b}=2Q_{3}Q_{4}=Q_{3}^{2}+Q_{4}^{2}-(Q_{3}-Q_{4})^{2}\,.\label{zb}\end{equation}
 Obviously, this factor cannot be rewritten in terms of the individual
charges $Q_{3,4}^{2}$ of the external particles.

When the Coulomb gauge is used for the on-shell amplitudes, DL contributions
are always proportional to $Q_{k}^{2}$ because they come from only
the self-energy graphs.\cite{sv,bw} Conversely, the interference
graphs shown in Fig.~1 (i.e., the graphs where the boson propagators
connect different external lines) do not yield DL contributions. This
feature makes the Coulomb gauge especially convenient for the calculations
of non-Abelian gauge theories; the nontrivial group factors are nothing
but the standard Casimir factors. On the contrary, when the Feynman
gauge is used, the DL contributions come from the interference graphs
shown in Fig.~1, whereas the self-energy graphs do not yield DL contributions.

While this simple pattern holds true for on-shell amplitudes, it becomes
more complex in the case of off-shell amplitudes. For off-shell amplitudes,
both the Feynman and Coulomb gauges received DL contributions from
the interference graphs. One can verify (using $\tilde{Q}^{2}=Z_{A}=Z_{B}=Z_{b}=2$)
that Eqs.~(\ref{solon}) and (\ref{solbig}) reproduce, as a particular
case, Eqs.~(\ref{sudon}) and (\ref{sudoff}) for the on-shell and
off-shell Sudakov form factors of the electron. Finally, the generalization
of these results to the case with more than 4 external particles can
be calculated in the DLA in a similar manner.

\section{Generalization to the EW Standard Model}

When the $2\rightarrow2$ scattering amplitudes of Sect.~2 are considered
in the framework of the ElectroWeak (EW) Standard Model, the situation
becomes more involved due to introduction of two cut-off scales, $\mu$
and $M$. We will only comment briefly on the on-shell EW amplitudes
in DLA, as these amplitudes have been computed in Refs.~\cite{flmm,b}
using the appropriate IREE's. We will examine in detail the $2\rightarrow2$
EW amplitudes in the hard kinematic limit.

In order to obtain the IREE for both photon and $W,Z$ exchanges,
one must factorize the DL contributions where the EW bosons have minimal
$k_{\perp}$. If the softest boson is a photon, then the lower limit
of integration over $k_{\perp}$ is $\mu$, and the result resembles
the QED case discussed in the previous section. Conversely, when the
softest boson is a $W$ or $Z$, then the lower limit of integration
over $k_{\perp}$ is $M$. Therefore, the integral on the RHS of Eq.~(\ref{intoffone})
corresponding to the factorization of the softest photon should be
replaced by the DL contributions of the factorization of the softest
$W$ and $Z$ bosons. The contributions of the softest $Z$ boson
looks quite similar to the photon case shown in Fig.~1; in contrast,
the $W$ boson exchange changes the flavors of the interacting fermions.

Considerable technical difficulties arise when using either the Feynman
or the Coulomb gauges.%
\footnote{For DL calculations with the Coulomb gauge, see e.g. Refs.~\cite{sv,bw}.%
} In the Coulomb gauge, the DL contributions of the softest bosons
come from the boson self-energy graphs, and this means that the amplitudes
are the exponential form obtained in Ref.~\cite{flmm}. On the contrary,
when the Feynman gauge is used, the factorization of the softest $W$-boson
leads to a system of four IREE's, the general solution of which consists
of four exponentials. If one of these four exponentials were negative,
this would lead to DL contributions that grow with $s$, and therefore
violate unitarity. However, the Born contributions (which are the
inhomogeneous terms in the integral IREE's) lead to the cancellation
of three of the exponentials. The fourth exponential then coincides
with the solution obtained using the Coulomb gauge. Consequently,
the Coulomb gauge is more convenient for calculations in the hard
kinematic limit of Eq.~(\ref{hard}), but becomes inappropriate when
working in the Regge kinematic region where the Feynman gauge is more
convenient. Indeed, in the Feynman gauge one can avoid the above technical
problems by expanding the EW amplitude into invariant ones, as was
shown in Ref.~\cite{b}.

\subsection{The Generalized IREE }

Let us consider a $2\rightarrow2$ electroweak scattering amplitude
$A(s,p_{3}^{2},p_{4}^{2},\mu^{2},M^{2})$ in the hard kinematic limit
of Eq.~(\ref{hard}). We use the notation of the previous Section
for the momenta of the external particles. We consider several cases
assuming again that the initial particles are always on-shell, whereas
the final particles can be either on-shell or off-shell. For any of
these cases, the amplitude obeys the following integral IREE: \begin{eqnarray}
A(s,p_{3}^{2},p_{4}^{2},\mu^{2},M^{2}) & = & A^{Born}-\frac{U}{16\pi^{2}}\int_{D}\frac{dk_{\perp}^{2}}{k_{\perp}^{2}}\frac{d\beta}{\beta}A(s,p_{3}^{2},p_{4}^{2},k_{\perp}^{2},M^{2})\nonumber \\
 &  & -\frac{g^{2}V}{16\pi^{2}}\int_{D'}\frac{dk_{\perp}^{2}}{k_{\perp}^{2}}\frac{d\beta}{\beta}A(s,p_{3}^{2},p_{4}^{2},k_{\perp}^{2},k_{\perp}^{2})\label{gen}\end{eqnarray}
 where $A^{Born}$ is the Born approximation for $A$, and $g=\sqrt{4\pi\alpha}/\sin\theta_{W}$
is the $SU(2)$-coupling of the EW Standard Model. The factors $U$
and $V$ depend on the particular relations between the cut-offs and
virtualities $p_{1,2}^{2}$; we will specify them below. The first
(second) integral in Eq.~(\ref{gen}) corresponds to the factorization
of the softest photon (weak boson). The integration regions $D$ and
$D'$ also depend on the relations between the parameters $p_{3}^{2},\, p_{4}^{2},\,\mu^{2},$
and $M^{2}$. Generally, the solution to Eq.~(\ref{gen}) can be
written as \begin{equation}
A=A^{Born}\exp(-\psi)\label{solgen}\end{equation}
 with appropriate exponents $\psi$. When both final-state particles
are on-shell, the integration over $\beta$ yields the common factor
$\ln(s/k_{\perp}^{2})$ in both the $U$ and $V$ integrals of Eq.~(\ref{gen})
even through the integration over $k_{\perp}^{2}$ runs from $\mu^{2}$
to $s$ in the first integral and from $M^{2}$ to $s$ in the second
one. In a manner similar to Eq.~(\ref{difon}), one obtains the on-shell
factor $U_{on}=32\pi^{2}\lambda Q^{2}$. The on-shell weak factor,
$C_{WZ}$, is the sum of the factors $V_{i}$: \begin{equation}
C_{WZ}=V_{1}+V_{2}+V_{3}+V_{4}\,,\label{von}\end{equation}
 where the subscripts 1 and 2 (3 and 4) refer to the initial-state
particles with momenta $p_{1,2}$ (the final-state particles with
momenta $p_{3,4}$). The factors $V_{i}$ are expressed in terms of
the Weinberg angle $\theta_{W}$, the weak isospins $T_{i}^{2}$ ($T_{i}^{2}=3/4$
for a fermion), the hypercharges $Y_{i}$, and the electric charges
$Q_{i}$ of the initial-state and final-state particles: \begin{equation}
V_{i}=[T_{i}^{2}+(Y_{i}^{2}/4)\tan^{2}\theta_{W}-Q_{i}^{2}sin^{2}\theta_{W}]\,.\label{vi}\end{equation}
 Differentiating Eq.~(\ref{gen}) with respect to the cut-offs, we
obtain the on-shell exponent $\psi\equiv\psi_{on}$: \cite{flmm}\begin{equation}
\psi_{on}=\alpha Q^{2}/(8\pi^{2})\ln^{2}(s/\mu^{2})+g^{2}/(32\pi^{2})C_{WZ}\ln^{2}(s/M^{2})\,.\label{psion}\end{equation}
 We recall that the photon double-logarithmic term $\ln^{2}(s/\mu^{2})$
in Eqs.~(\ref{solon},\ref{psion}) is obtained under the assumption
that the cut-off $\mu$ is greater than any mass involved.

In the case where the initial-state particles are light but the final-state
particles are so heavy that Eq.~(\ref{musmall}) is satisfied, the
term $Q^{2}\ln^{2}(s/\mu^{2})$ should be replaced by \begin{equation}
Q^{2}\ln^{2}(s/\mu^{2})-(Z_{A}/2)\ln^{2}(m_{1}^{2}/\mu^{2})-(Z_{B}/2)\ln^{2}(m_{2}^{2}/\mu^{2})\label{smallmu}\end{equation}
 in agreement with Eq.~(\ref{solmusmallexpl}).

\subsection{The off-shell case:}

When the final-state particles are off-shell, the regions $D$ and
$D'$ in Eq.~(\ref{gen}) are more complicated. We study below several
interesting situations, denoting these as $R_{1,2,3,4}$.

\subsubsection{$R_{1}$ Case: final-state particles of small virtuality}

We call $R_{1}$ the simplest case when the virtualities of the final
particles are relatively small: \begin{eqnarray}
\mu^{2} & < & p_{3,4}^{2}<M^{2}\ll s\,,\nonumber \\
p_{3}^{2}\, p_{4}^{2} & < & s\mu^{2}\qquad.\label{r1}\end{eqnarray}
 This case is quite similar to Eq.~(\ref{smallvirt}), so we immediately
conclude that the exponent $\psi_{1}$ of Eq.~(\ref{solgen}) for
this case is \begin{equation}
\psi_{1}=\frac{\alpha}{8\pi}[Q^{2}\ln^{2}(s/\mu^{2})-Z_{A}\ln^{2}(p_{3}^{2}/\mu^{2})-Z_{B}\ln^{2}(p_{4}^{2}/\mu^{2})]+\frac{g^{2}}{32\pi^{2}}C_{WZ}\ln^{2}(s/M^{2})\,.\label{psi1}\end{equation}

\subsubsection{$R_{2}$ Case: Deeply virtual for photon, and on-shell for $W/Z$
boson}

In the second case, $R_{2}$, the virtualities $p_{3,4}^{2}$ are
larger and satisfy \begin{eqnarray}
\mu^{2} & < & p_{3,4}^{2}<M^{2}\,,\nonumber \\
s\mu^{2} & < & p_{3}^{2}\, p_{4}^{2}<sM^{2}\qquad.\label{r2}\end{eqnarray}
 This kinematics are deeply-virtual for the softest photon (with the
solution given by Eq.~(\ref{solbig})), but at the same time it is
on-shell for the softest $W,Z$-bosons (with the solution given by
Eq.~(\ref{solon})). Therefore the exponent $\psi_{2}$ for this
case is \begin{equation}
\psi_{2}=\lambda[\tilde{Q}^{2}x^{2}-\tilde{Z}_{A}z_{1}^{2}-\tilde{Z}_{B}z_{2}^{2}+2Z_{b}(x-z_{1})(x-z_{2})]+\frac{g^{2}}{32\pi^{2}}C_{WZ}\ln^{2}(s/M^{2})~\label{psi2}\end{equation}

\subsubsection{$R_{3}$ Case: Deeply virtual for photon, and moderately virtual
for $W/Z$ boson}

The case $R_{3}$, defined as \begin{eqnarray}
p_{3,4}^{2} & > & M^{2}\nonumber \\
s & < & p_{3}^{2}\, p_{4}^{2}<s\, M^{2}\quad,\label{r3}\end{eqnarray}
 describes a situation which is deeply-virtual for the photon, and
moderately-virtual for the $W,Z$-bosons. Combining the results of
Eq.~(\ref{solsmall}) and Eq.~(\ref{solbig}), we find the solution:\begin{eqnarray}
\psi_{3} & = & \lambda[\widetilde{Q}^{2}x^{2}-\widetilde{Z}_{A}z_{1}^{2}-\widetilde{Z}_{B}z_{2}^{2}+2Z_{b}(x-z_{1})(x-z_{2})]\label{psi3}\\
 &  & +\frac{g^{2}}{32\pi^{2}}\big[C_{WZ}\ln^{2}(s/M^{2})-2V_{3}\ln^{2}(p_{3}^{2}/M^{2})-2V_{4}\ln^{2}(p_{4}^{2}/M^{2})\big]\,.\nonumber \end{eqnarray}

\subsubsection{$R_{4}$ Case: Deeply virtual for photon $W/Z$ boson}

Finally, when the momentum $p_{3,4}^{2}$ are so large that \begin{eqnarray}
p_{3,4}^{2} & > & M^{2}\nonumber \\
p_{3}^{2}\, p_{4}^{2} & > & sM^{2}\,,\label{r4}\end{eqnarray}
 the situation is deeply-virtual for all of the softest electroweak
bosons, and this case is equivalent to the scheme with a single cut-off
$M$. The EW exponent $\psi_{4}$ for this case corresponds to the
Eq.~(\ref{solbig}): \begin{eqnarray}
\psi_{4} & = & \big[\lambda\widetilde{Q}^{2}\ln^{2}(s/\mu^{2})+g^{2}\widetilde{C}_{WZ}/(32\pi^{2})\ln^{2}(s/M^{2})\big]\nonumber \\
 & - & \big[\lambda\widetilde{Z}_{A}\ln^{2}(p_{3}^{2}/\mu^{2})+g^{2}\widetilde{X}_{3}/(32\pi^{2})\ln^{2}(p_{3}^{2}/M^{2})\big]\nonumber \\
 & - & \big[\lambda\widetilde{Z}_{B}\ln^{2}(p_{4}^{2}/\mu^{2})+g^{2}\widetilde{X}_{4}/(32\pi^{2})\ln^{2}(p_{4}^{2}/M^{2})\big]\nonumber \\
 & + & 2\big[\lambda Z_{b}+g^{2}X_{b}/(32\pi^{2})\big]\ln^{2}(s/p_{3}^{2})\ln^{2}(s/p_{4}^{2})\quad.\label{psi4}\end{eqnarray}
 Here, $X_{3,4}=2V_{3,4}-X_{b}$, and $X_{b}=1/2+(1/\cos^{2}\theta_{W})[t_{3}^{(3)}-Q_{3}\sin^{2}\theta_{W}][t_{3}^{(4)}-Q_{4}\sin^{2}\theta_{W}]$,
where $Q_{3,4}$ are the electric charges and $t_{3}^{3,4}$ are the
eigenvalues of the $SU(2)$ -generator $T_{3}$ acting on the final-state
particles 3 and 4, respectively. It is convenient to rewrite $\psi_{4}$
in the following form: \begin{eqnarray}
\psi_{4} & = & \big[\lambda Q^{2}\ln^{2}(s/\mu^{2})+g^{2}C_{WZ}/(32\pi^{2})\ln^{2}(s/M^{2})\big]\nonumber \\
 & - & \big[\lambda Z_{A}\ln^{2}(p_{3}^{2}/\mu^{2})+g^{2}X_{3}/(32\pi^{2})\ln^{2}(p_{3}^{2}/M^{2})\big]\nonumber \\
 & - & \big[\lambda\widetilde{Z}_{B}\ln^{2}(p_{4}^{2}/\mu^{2})+g^{2}X_{4}/(32\pi^{2})\ln^{2}(p_{4}^{2}/M^{2})\big]\nonumber \\
 & + & \big[\lambda Z_{b}+g^{2}X_{b}/(32\pi^{2})\big]\ln^{2}(sM^{2}/p_{3}^{2}p_{4}^{2})\quad.\label{psi4a}\end{eqnarray}
 Eq.~(\ref{solgen}), together with the phases specified in Eqs.~(\ref{psi1},\ref{psi2},\ref{psi3},\ref{psi4},\ref{psi4a}),
describe the $2\rightarrow2$ off-shell electroweak scattering amplitudes
in the hard kinematic limit for several virtualities of the final-state
particles. The above techniques can be generalized for other particular
cases in a similar manner.

\section{Off-shell amplitudes in the Regge kinematics}

The evaluation of the DL contributions to the on-shell and off-shell
scattering amplitudes in the hard kinematic limit (\ref{hard}) is
relatively straightforward, and leads to the exponentiation of the
one-loop DL contributions. When the scattering amplitudes are evaluated
in the forward Regge limit \begin{equation}
s\approx-u\gg-t\quad,\label{forward}\end{equation}
 or backward Regge limit\begin{equation}
s\approx-t\gg-u\quad,\label{backward}\end{equation}
 the DL expressions are more complicated.

The $2\rightarrow2$ EW Regge amplitudes at TeV energies for electron-positron
colliders have been discussed previously in Refs.~\cite{b}. In the
following, we will consider one example of the off-shell Regge amplitudes
in the QCD theory. We obtain a relation between on-shell and off-shell
amplitudes for quark scattering, and use this result to calculate
the non-singlet structure functions for Deep Inelastic Scattering
when the initial quark is off-shell. In doing so, we make use of the
expressions for the on-shell non-singlet structure functions obtained
in Ref.~\cite{egt1} at small $x$. These expressions account for
the DL and single-logarithmic (SL) contributions to all orders in
the QCD coupling $\alpha_{s}$. Explicit expressions combining the
DGLAP\cite{dglap} expressions and our total resummation (DL and SL
contributions) are obtained in Ref.~\cite{egt3}. These combined
results are especially important in the region of small $x$, as the
combination leads to the power-like asymptotic behavior $(\sim x^{-\omega_{0}})$
of the structure functions when $x\rightarrow0$, with intercept $\omega_{0}\approx0.4$.
We briefly summarize these on-shell results.

According to the Optical Theorem, it is possible to relate the on-shell
non-singlet structure function $F^{(\pm)}(x,Q^{2})$ to the imaginary
part of the forward Compton amplitude $A^{(\pm)}(x,Q^{2})$ with $-t\leq\mu^{2}$.
It is convenient to present $A^{(\pm)}(x,Q^{2})$ in the form of the
Mellin integral: \begin{equation}
A^{(\pm)}(x,Q^{2})=\int_{-\imath\infty}^{+\imath\infty}\frac{d\omega}{2\pi\imath}\Big(\frac{s}{\mu^{2}}\Big)^{\omega}\xi^{(\pm)}(\omega)T^{(\pm)}(\omega,Q^{2})\,,\label{mellin}\end{equation}
 with the signature factor $\xi^{(\pm)}(\omega)=-(e^{-\imath\pi\omega}\pm1)/2$,
total energy $s=Q^{2}/x$, and the infrared cut-off $\mu$. The Mellin
amplitude $T^{(\pm)}(\omega,Q^{2})$ obeys the following IREE: \begin{equation}
\partial T^{(\pm)}/\partial y+\omega T^{(\pm)}=T^{(\pm)}H^{(\pm)}(\omega)\,,\label{eqt}\end{equation}
 with anomalous dimensions $H^{(\pm)}$, and $y=\ln(Q^{2}/\mu^{2})$
so that $\mu^{2}$ is the starting point of the $Q^{2}$-evolution.
$H^{(\pm)}$ include both double-logarithmic and the most important
part of the single-logarithmic contributions. From Refs.~\cite{egt1}
we have \begin{equation}
H^{(\pm)}=(1/2)\Big[\omega-\sqrt{\omega^{2}-B^{\pm}(\omega)}\Big]\label{hns}\end{equation}
 where \begin{equation}
B^{(\pm)}(\omega)=[(1+\omega/2)4\pi C_{F}A(\omega)+D^{(\pm)}(\omega)]/(2\pi^{2})\,.\label{b}\end{equation}
 In Eq.~(\ref{b}), $D^{(\pm)}(\omega)$ and $A(\omega)$ can be
expressed in terms of $\rho=\ln(1/x)$, $b=(33-2n_{f})/12\pi$, and
the color factors $C_{F}=4/3$, $N=3$: \begin{equation}
D^{(\pm)}(\omega)=\frac{2C_{F}}{b^{2}N}\int_{0}^{\infty}d\eta e^{-\omega\eta}\ln\big(\frac{\rho+\eta}{\eta}\big)\Big[\frac{\rho+\eta}{(\rho+\eta)^{2}+\pi^{2}}\mp\frac{1}{\eta}\Big]~,\label{d}\end{equation}
\begin{equation}
A(\omega)=\frac{1}{b}\Big[\frac{\eta}{\eta^{2}+\pi^{2}}-\int_{0}^{\infty}\frac{d\rho e^{-\omega\rho}}{(\rho+\eta)^{2}+\pi^{2}}\Big]\,.\label{a}\end{equation}
 In Eq.~(\ref{a}), $A(\omega)$ is the Mellin transform of $\alpha_{s}(k^{2})=1/(b\,\ln(-k^{2}/\Lambda^{2}))$
with time-like argument $k^{2}$. Solving Eq.~(\ref{eqt}) and introducing
$F^{(\pm)}(x,Q^{2})$, \begin{equation}
F^{(\pm)}(x,Q^{2})=\frac{1}{\pi}\Im A^{(\pm)}(x,Q^{2})\,,\label{f}\end{equation}
 we obtain\begin{equation}
F^{(\pm)}(x,Q^{2})=(e_{q}^{2}/2)\int_{-\imath\infty}^{+\imath\infty}\frac{d\omega}{2\pi\imath}(1/x)^{\omega}C^{(\pm)}(\omega)\delta q(\omega)\exp\big(H^{(\pm)}(\omega)y\big)~.\label{fon}\end{equation}
 Eq.~(\ref{f}) relates $F^{(+)}$ and the forward amplitude $A^{(+)}$
via the Optical theorem. In fact, $F^{(+)}$ is the non-singlet contribution
to the structure function $F_{1}$, and $F^{(-)}$ is the non-singlet
contribution to the polarized structure function $g_{1}$. $\delta q$
is the initial quark density, which is commonly determined from fitting
the experimental data. The coefficient functions $C^{(\pm)}$ can
be expressed in terms of the anomalous dimensions $H^{(\pm)}$ as:
\begin{equation}
C^{(\pm)}=\frac{\omega}{\omega-H^{(\pm)}(\omega)}\,.\label{cns}\end{equation}

We now consider the case where the initial quark is off-shell with
virtuality $p^{2}$,%
\footnote{As in the previous Sections, we drop the modulus sign in Eq.~(\ref{virtq}),
keeping the notation $p^{2}$ instead of $|p^{2}|$.%
}\begin{equation}
\mu^{2}<|p^{2}|\ll Q^{2}\,.\label{virtq}\end{equation}
 When the initial quark is off-shell, the forward Compton amplitude
$\widetilde{A}^{(\pm)}$ depends on $p^{2}$ as well: $\widetilde{A}^{(\pm)}=\widetilde{A}^{(\pm)}(s,Q^{2},p^{2},\mu^{2})$.
As before, it is convenient to work with the Mellin off-shell amplitude
$G(\omega,Q^{2},p^{2},\mu^{2})$ which is related to $\widetilde{A}^{(\pm)}(s,Q^{2},p^{2},\mu^{2})$
through the Mellin transform (\ref{mellin}). The IREE for the off-shell
amplitude $G(\omega,Q^{2},p^{2},\mu^{2})$ is similar to the one for
the on-shell amplitude $T^{(\pm)}$: \begin{equation}
\partial G^{(\pm)}(\omega,y,z)/\partial y+\partial G^{(\pm)}(\omega,y,z)/\partial z+\omega G^{(\pm)}(\omega,y,z)=T^{(\pm)}(\omega,y)h^{(\pm)}(\omega,z)\label{eqoff}\end{equation}
 where $z=\ln(p^{2}/\mu^{2})$. $h^{(\pm)}(\omega,z)$ are new anomalous
dimensions which can be found from the following IREE: \begin{equation}
\partial h^{(\pm)}/\partial z+\omega h^{(\pm)}=D^{(\pm)}+h^{(\pm)}H^{(\pm)}(\omega)\,,\label{eqh}\end{equation}
 with $D^{(\pm)}$ defined in Eq.~(\ref{b}).

Solving Eq.~(\ref{eqh}), and using the matching conditions for $z=0$

\begin{equation}
h^{(\pm)}(\omega,z)|_{z=0}=H^{(\pm)}(\omega)\label{matchh}\end{equation}
 we obtain

\begin{equation}
h^{(\pm)}=\Big[H^{(\pm)}+\int_{0}^{z}du\, e^{(\omega-H^{(\pm)})u}D^{(\pm)}(u)\Big]e^{-(\omega-H^{(\pm)})z}\,.\label{solh}\end{equation}
 Once the $h^{(\pm)}$ are known, we can calculate $G^{(\pm)}$. Substituting
$h^{(\pm)}$ and $T^{(\pm)}$ into Eq.~(\ref{eqoff}) and solving,
we obtain

\begin{equation}
G^{(\pm)}(\omega,y,z)=T^{(\pm)}(\omega,2\eta)e^{-\omega z}+e^{-\omega l}\int_{\eta}^{l}du\, e^{\omega u}T^{(\pm)}(\omega,u+\eta)h^{(\pm)}(u-\eta)\label{solg}\end{equation}
 where $l=(y+z)/2,\eta=(y-z)/2$. Eq.~(\ref{solg}) gives $G^{(\pm)}(\omega,y,z)=T^{(\pm)}(\omega,y)$
when $z=0$.

To demonstrate a simple application of Eq.~(\ref{solg}), let us
estimate the effect of the virtuality $p^{2}$ of the initial quark
on the small-$x$ asymptotics of the off-shell non-singlet structure
functions $\widetilde{F}^{(\pm)}$. Given the results of Refs.~\cite{egt1},
when $x\rightarrow0$ the small-$x$ asymptotics of the on-shell structure
functions $F^{(\pm)}$ are Regge-like: \begin{equation}
F^{(\pm)}(x,Q^{2})\sim e_{q}^{2}\delta q(\omega_{0}^{(\pm)})c^{(\pm)}\xi^{\omega_{0}^{(\pm)}}/\ln^{3/2}\xi\label{asympton}\end{equation}
 where $\xi=\sqrt{Q^{2}/(x^{2}\mu^{2})}$, $c^{(\pm)}=\Big[2(1-q)/(\pi\sqrt{B^{(\pm)}})\Big]^{1/2}$,
$q^{(\pm)}=d\sqrt{B^{(\pm)}}/d\omega|_{\omega=\omega_{0}^{(\pm)}}$,
and the intercepts are $\omega_{0}^{(+)}=0.38$ and $\omega_{0}^{(-)}=0.43$.
To estimate the asymptotics of $\widetilde{F}^{(\pm)}(x,Q^{2},p^{2})$,
we neglect $D^{(\pm)}$ in Eq.~(\ref{solh}) since the effect of
$D^{(\pm)}$ on the small-$x$ behavior of $g_{1}$ is not large.\cite{egt1}
Performing the integration in Eq.~(\ref{solg}) using the saddle-point
method, we obtain at the following expression for the small-$x$ asymptotics
of the off-shell structure functions $\widetilde{F}^{(\pm)}$:

\begin{equation}
\widetilde{F}^{(\pm)}(x,Q^{2},p^{2})\approx F^{(\pm)}(x,Q^{2})e^{-\omega_{0}z/2}\Big[1+\omega_{0}^{(\pm)}z/2\Big]\,,\label{asymptonoff}\end{equation}
 with $\omega_{0}^{(+)}=0.38$ and $\omega_{0}^{(-)}=0.43$.%
\footnote{See Refs.~\cite{egt1} for details.%
} In order to estimate the difference between the on-shell and the
off-shell non-singlet structure functions, we define the deviation
$R$: \begin{equation}
R=\big(F^{(\pm)}-\widetilde{F}^{(\pm)}\big)/F^{(\pm)}\approx1-e^{-0.2\ln(p^{2}/\mu^{2})}\Big[1+0.2\ln(p^{2}/\mu^{2})\Big]\,,\label{R}\end{equation}
 where we estimate $\mu$ to be $\approx1$~GeV.\cite{egt1} We observe
from Eq.~(\ref{R}) that $R$ increases with increasing $p^{2}$.
By definition, $R=0$ when $p^{2}=1$~GeV$^{2}$ and therefore it
grows when $p^{2}$ becomes greater than $\mu^{2}$ (we remind that
the notation $p^{2}$ actually denotes $|p^{2}|$ ). In particular,
$R\approx0.02$ for $|p^{2}|=3$~GeV$^{2}$ and $R\approx0.08$ for
$|p^{2}|=10$~GeV$^{2}$.

\section{Conclusions}

In this paper, we have presented an approach for calculating the off-shell
scattering amplitudes in the double-logarithmic (DL) approximation.
We have considered, in particular, the case where the initial particles
are on-shell, but some of the final particles can be off-shell. Technically,
this approach is based on constructing the appropriate infrared evolution
equations (IREE) for the off-shell amplitudes. The specific form of
the solutions depends on the size of the virtualities of the final
particles. The general strategy for obtaining the off-shell solutions
was to use the matching conditions to relate amplitudes of the same
process, but with smaller virtualities of the final particles. Such
a procedure allows us to relate the amplitudes with $N$ off-shell
particles to those with $N-1$ off-shell particles. Using this recursive
relation, we can eventually relate the off-shell amplitudes to the
on-shell amplitudes.

To demonstrate the utility of this approach, we have computed the
scattering amplitudes in the hard kinematic limit of Eq.~(\ref{hard}),
where the DL corrections simply exponentiation the one-loop DL contribution.
We have reproduced the well-known results for the $2\rightarrow2$
on-shell QED amplitudes of Eqs.~(\ref{solon},\ref{solmusmallexpl}),
and used these results to obtain explicit expressions for the amplitudes
with one (Eq.~(\ref{solone})) and two (Eqs.~(\ref{solsmall},\ref{solbig}))
final particles off-shell. We have considered both cases of moderate
and large virtualities for the final particles, and compared the results
in the Feynman and Coulomb gauges. In particular, we have shown that
both the self-energy and interference graphs yield DL contributions
in the Coulomb gauge; this result is contrary to the on-shell case.
This result leads to the conclusion that the Coulomb gauge does not
have provide any technical advantage, compared to the Feynman gauge,
for calculating the off-shell amplitudes.

In Sect.~3 we have computed the more complex case of the off-shell
amplitudes for the $2\rightarrow2$ electroweak processes in the hard
kinematic limit at energies $\gg100$~GeV. In addition to the virtual
photon exchanges, we also took the exchanges of virtual $W$ and $Z$
bosons into account. The results of the DL contributions for these
scattering amplitudes depends on the virtualities of the final particles;
the results are presented in Eqs.~(\ref{psi1},\ref{psi2},\ref{psi3},\ref{psi4}).

Finally, we have studied the QCD forward Compton scattering amplitude
in the Regge kinematic limit (Eqs.~(\ref{forward},\ref{backward}))
with both the photons and the quarks being off-shell. This result
was used to calculate the DIS non-singlet structure functions with
off-shell initial quarks (Eq.~(\ref{solg})). To estimate the difference
between the off-shell and the on-shell structure functions, we compared
their small-$x$ asymptotics (Eqs.~(\ref{asympton},\ref{asymptonoff})),
and the result is presented in Eq.~(\ref{R}). The on-shell and off-shell
asymptotic amplitudes differ at larger values of $p^{2}$; for example,
this difference is $\approx8\%$ for $p^{2}=10\, GeV^{2}$. A more
accurate estimate of the effect of the virtuality on the structure
functions could be directly obtained from the explicit expression
of Eq.~(\ref{solg}), instead of using their asymptotic behavior
of Eq.~(\ref{asymptonoff}); this topic will be reserved for future
study.

\section{Acknowledgment}

We are grateful to A.~Barroso for useful discussions. The work is
supported in part by grant RSGSS-1124.2003.2, U.S. DoE grant DE-FG03-95ER40908,
and the Lightner-Sams Foundation.


\begin{thebibliography}{10}
\bibitem{dglap}G.~Altarelli and G.~Parisi, Nucl.~Phys.B126 (1977) 297; V.N.~Gribov
and L.N.~Lipatov, Sov.~J.~Nucl.~Phys. 15 (1972) 438; L.N.Lipatov,
Sov.~J.~Nucl.~Phys. 20 (1972) 95; Yu.L.~Dokshitzer, Sov.~Phys.~JETP
46 (1977) 641. 
\bibitem{sud}V.V.~Sudakov. Sov. Phys. JETP 3(1956)65. 
\bibitem{et}B.I.~Ermolaev and S.I.~Troyan. Nucl. Phys.B590(2000)521. 
\bibitem{e}B.I.~Ermolaev. Yad. Fiz. 28(1085)1978; P.~Ciafaloni and D.~Comelli.
Phys.Lett.B 476(2000)49; M.~Hori, H.~Kawamura and J.~Kodaira. Phys.Lett.B
491(2000)275; M.~Melles. Phys.Lett.B 495(200)81; A.~Denner, M.~Melles
and S.~Pozzorini. Nucl.Phys.B 662(2003)299. 
\bibitem{bw}W.~Beenakker and A.~Werthenbach. Nucl.Phys.B 630(2002)275; 
\bibitem{flmm}V.S. Fadin, L.N. Lipatov, A.D.~Martin, and M.~Melles, Phys.Rev.
D61 (2000) 094002. 
\bibitem{ciaf}M.~Ciafaloni, P.~ Ciafaloni, D.~Comelli. Phys.Rev.Lett.88(2002)102001. 
\bibitem{k}J.H.~Kuhn, A.A.~Penin. hep-ph/9906545; J.H.~Kuhn, A.A.~Penin,
and V.A.~Smirnov. Eur. Phys. J.C. 17(2000)97. 
\bibitem{kur}E. Bartos, E.A. Kuraev, I.O. Cherednikov. Phys.Lett.B593(2004)115. 
\bibitem{b}B.I.~Ermolaev, M.~Greco and S.I.~Troyan. Phys.Rev.D67(014017)2003;
B.I.~Ermolaev, S.M.~Oliveira and S.I.~Troyan. Phys.Rev.D66(114018)2002;
A.~Barroso, B.I.~Ermolaev, M.~Greco S.M.~Oliveira, and S.I.~Troyan.
Phys.Rev.D69(034012)2004. 
\bibitem{g}V.N.~Gribov. Yad. Fiz. 5(1967)199. 
\bibitem{kl}R.~Kirschner and L.N.~Lipatov. Nucl. Phys. B213(1983)122 
\bibitem{efl}B.I~Ermolaev, V.S.~Fadin, L.N.~Lipatov. Yad. Fiz. 45(1987)817;
B.I.~Ermolaev. Sov. J. Nucl. Phys. 47(841)1988. 
\bibitem{ce}M.~Chaichian and B.~Ermolaev. Nucl. Phys.B451(1995)194. 
\bibitem{egt1}B.I.~Ermolaev, M.~Greco and S.I.~Troyan. Nucl.Phys.B 594 (2001)71;
ibid 571(2000)137; Phys.Lett.B579(321),2004. 
\bibitem{sv}A.V.~Smilga and M.I.~Vysotsky. Nucl. Phys. B150(1979)173. 
\bibitem{egt3}B.I.~Ermolaev, M.~Greco and S.I.~Troyan. hep-ph/0503019. \end{thebibliography}
\end{document}